# Hafnia-based Double Layer Ferroelectric Tunnel Junctions as Artificial Synapses for Neuromorphic Computing

*Benjamin Max\*[1], Michael Hoffmann[2], Halid Mulaosmanovic[2], Stefan Slesazeck[2], and Thomas Mikolajick[1,2]*

[1] Technische Universität Dresden, Chair for Nanoelectronics, Institute for Semiconductors and Microsystems, 01069 Dresden, Germany

[2] NaMLab gGmbH, Noethnitzer Straße 64 a, 01187 Dresden, Germany

ABSTRACT

Ferroelectric tunnel junctions (FTJ) based on hafnium zirconium oxide ($Hf_{1-x}Zr_xO_2$; HZO) are a promising candidate for future applications, such as low-power memories and neuromorphic computing. The tunneling electroresistance (TER) is tunable through the polarization state of the HZO film. To circumvent the challenge of fabricating thin ferroelectric HZO layers in the tunneling range of 1-3 nm range, ferroelectric/dielectric double layer sandwiched between two symmetric metal electrodes are used. Due to the decoupling of the ferroelectric polarization storage layer and a dielectric tunneling layer with a higher bandgap, a significant TER ratio between the two polarization states is obtained. By exploiting previously reported switching behaviour and the gradual tunability of the resistance, FTJs can be used as potential candidates for the emulation of synapses for neuromorphic computing in spiking neural networks. The implementation of two major components of a synapse are shown: long term depression/potentiation by varying the amplitude/width/number of voltage pulses applied to the artificial FTJ synapse, and spike-timing-dependent-plasticity curves by applying time-delayed voltages at each electrode. These experimental findings show the potential of spiking neural networks and neuromorphic computing that can be implemented with hafnia-based FTJs.

INTRODUCTION

In recent years, a great effort has been made on mimicking the functionality of the human brain with electronic circuits and artificial neural networks.[1,2] Especially the ability for efficient learning and computing big data was investigated.[3] A large number of biologically inspired computing and neuromorphic systems have been proposed.[4-6] However, many state-of-the-art applications are suffering from applying the analogue computing mechanism in the brain to the digital von-Neumann architecture that is prevalent in commonly available computer systems. Drawbacks in these implementations are high power consumption, limited parallel processing resulting in long computational times and scalability problems to large networks.[7,8] New emerging memory and logic devices are promising candidates for the implementation in new circuits and designs that describe spiking neural networks (SNN).[9,10] Various devices have been proposed, such as resistive memories (RRAM),[11-13] phase-change memories (PCM),[14,15] or ferroelectric devices.[16-19] A common feature of these device is the programmable resistance that is used for the learning state – representing the synaptic weight - of the artificial synapse cell. The working mechanism of a SNN crossbar array based on these devices is the gradual switching from a high resistance state (HRS) to a lower resistance state (LRS) and vice versa under the application of electric fields. If the device exhibits multilevel resistance states between the minimum and maximum values, an emulated learning process through synaptic plasticity can be achieved. Precise reproducibility and non-volatile retention of the resistance states, especially in regard to Hebbian learning and memory, is crucial for implementation in SNNs.[20,21] In this paper we show that ferroelectric tunnel junctions (FTJ) can be used to mimic the behavior of a synaptic cell, and are therefore suitable candidates for implementation in SNNs.

A two-terminal FTJ consists of a ferroelectric layer that is sandwiched between two electrodes. The reversal of the polarization state leads to different band alignments where the effective tunneling barrier height and width is changed.[22] Therefore a different tunneling coefficient is obtained when an electric field is applied at the electrodes, resulting in two different tunneling electroresistances (LRS and HRS) for each polarization state. Ferroelectric tunnel junctions based on hafnium zirconium oxide ($Hf_{1-x}Zr_xO_2$; HZO) offer several advantages compared to other devices, such as CMOS-compatible manufacturing processes, low power consumption, fast switching speeds and long data retention.[23-25] A 50:50 stoichiometry between hafnium and zirconium has been shown to exhibit excellent ferroelectric properties with remanent polarization values up to 25 µC cm$^{-2}$,[26] with a comparably low thermal budget[27] and good retention and wakeup/fatigue behavior. Using the classical approach of metal-ferroelectric-metal FTJs, the desired thickness of the ferroelectric film must be in the low nanometer range (1-3 nm HZO thickness) to enable sufficiently high tunneling currents.[28,29] Film deposition for these low thicknesses while maintaining acceptable ferroelectric properties has been shown to be extremely difficult.[30-32] If the $2P_r$ value between the polarization states of the ferroelectric film is too low, no significant difference in the tunneling electroresistance (TER) ratio is achievable. Moreover, such thin films suffer from high background leakage currents through the crystalline film that will not allow high memory windows for gradually tuning the electroresistance state.[32] We have recently proposed and demonstrated a two-layer ferroelectric tunnel junction that consists of symmetric titanium nitride electrodes and a hafnium zirconium oxide/aluminum oxide double-layer stack in between (TiN/$Hf_{0.5}Zr_{0.5}O_2$/$Al_2O_3$/TiN).[33,34] By decoupling the polarization storage layer (HZO) and the tunneling barrier layer ($Al_2O_3$) we have shown a TER ratio of 10 and good endurance and retention behavior of these FTJs. Utilizing metal electrodes allows an easier integration into back end of line of CMOS processes, which is a significant advantage compared to previously reported hafnia-based FTJs for neuromorphic computing.[19] The silicon electrode that is used in those devices is only available at the substrate, which limits their use to only a single layer approach in the front end of line. Additionally, the semiconductor electrode introduces an additional depolarization field, which is detrimental to the retention behavior of these devices. In this paper, we utilize our devices for artificial synapses and show potentiation and depression characteristics through the variation of pulse amplitude, pulse width and number of pulses. Spike-timing-dependent-plasticity (STDP) can be verified by applying time-delayed voltage pulses at top and bottom electrodes.

## RESULTS AND DISCUSSION

SNNs emulate biological neuronal systems which communicate by generating and propagating electrical pulses (action potentials). In a very simple SNN, the artificial neurons are connected via electrical synapses and transport information through sequences of spikes where neural information is carried by the timing of spikes rather than the specific shape.[35,36] The neuron can be modeled through various means, e.g. (leaky)-integrate-and-fire neuron models.[16,37] The neurons connect and communicate via synapses. An input spike from the pre-neuron at the synapse leads to an input signal into the postsynaptic neuron. The connecting synapse is responsible for transmitting and modulating that input signal. Storing information and learning in synapses is realized through changing the synaptic plasticity. There are various forms of changing this synaptic weight which are mostly relevant on different timescales. Short-term potentiation and depression (STP/STD) are relevant in the milliseconds to minutes range, while long-term potentiation and depression (LTP/LTD) characteristics persist for days or longer.[38] The synaptic plasticity can also change depending on the temporal order and timing between the incoming pre-neuron action potential and the post-neuron action potential (STDP). The latter is commonly observed in unsupervised Hebbian learning, where the synaptic strength is modulated by repeated and persistent stimulation and its time correlation.[39] Applications for STDP-like processes can e.g. be found in cluster analysis and pattern recognition.[40,41]

In the following part we will show that hafnia-based FTJs can be used for mimicking these synapses by experimentally verifying the previously mentioned synaptic elements. To understand the neuromorphic characteristics, first the switching mechanisms of the TiN/HZO/Al$_2$O$_3$/TiN FTJ are investigated. The switching behavior can be seen in **figure 1**. The schematic device structure is shown in figure 1d. More information and details about the basic operation of these devices can be found elsewhere.[33,34] The ferroelectric layer in the FTJ can be set in two different polarization states, resulting in different band tilting and electric potential distributions across the HZO and aluminum oxide layer. The asymmetric stack structure with the aluminum oxide leads to an imperfect screening of the polarization charges and results in an electric field across the ferroelectric and dielectric films. The tunneling current forced by an externally applied bias is dependent on the tunneling width and barrier height of the $HZO/Al_2O_3$ double layer. The two different electric potentials lead to switching between tunneling through the thin dielectric layer only (LRS) and modified Fowler-Nordheim tunneling trough the dielectric and parts of the ferroelectric (HRS). This results in a strong difference in the tunneling current for the two polarization states. The tunability of the overall resistance state with different voltage pulse amplitudes is shown in figure 1a. The HRS at 14 G$\Omega$ can be obtained by applying a negative voltage to the bottom electrode, thus resetting the device into the $P_{down}$ state leading to a low tunneling current. The set operation with positive amplitudes in the range of 4-8 V leads to a higher tunneling current and the LRS at approximately 1 G$\Omega$ is reached. The resistance hysteresis loop for the FTJ is a direct consequence of the polarization reversal of the HZO layer.[34] For these comparably large devices (200μm diameter) the absolute tunneling current that we observe is already rather low. When scaling down the FTJ devices to a diameter in the range of 100 nm to be adopted with state-of-the-art CMOS technology nodes, the current would decrease by approximately six orders of magnitudes. However, recently it has been shown that FTJ devices with an area of just 200 nm x 200 nm can be read out when adopting a suitable readout circuit.[42] Therefore, further increasing the current density, e.g. through further stack optimization regarding the HZO and dielectric thickness, better control of the deposition processes or using a different dielectric barrier material, has to be considered for integration into circuits or crossbar neural networks. One should also note that for parallel processing in large arrays which are necessary for massive parallel computing, a small current is one of the key ingredients to minimize the total power consumption and to decrease disturbance through sneak path currents and parasitic voltage in the array.[43,44] Longer sensing times may be the consequence which can be tolerated in the application envisioned here in contrast to high speed random access memories were fast sensing is a must. Therefore, the ultra-low current device is a good starting point with respect to the target application discussed here. It is possible to set multiple intermediate states with good reproducibility over 50 cycles, shown in figure 1b. These multiple levels show some retention loss when extrapolated to 10 years (see figure 1c). While the states are clearly separable during initial measurements, the memory window (defined as the difference between highest and lowest resistance) declines over time due to the inherent depolarization fields, leading to a reduction of the stored remanent polarization. We have previously shown that by using built-in bias fields (implemented by different metal electrodes with different work functions such as titanium nitride and platinum) the polarization states can be stabilized to maintain the memory window.[45,46] The improvement regarding the retention is important also for intermediate states, where decrease of the memory window would make the readout operation of these in-between current states even more difficult. Note here that retention measurements were done at room temperature. The memory window will be further reduced when higher temperatures are considered. The above mentioned bias fields together with additional tuning of the field distribution in the device by adjusting the material and thickness of the tunneling layer and possibly some further improvement of the ferroelectric HZO film towards higher coercivity can improve the retention characteristics.[45,46] Moreover, in FeFETs that use a similar stack, but have an additional depolarization component from the $SiO_2$ interface layer and semiconducting channel, excellent retention has previously been demonstrated even at high temperatures.[47] This optimization is essential for keeping the learning effects with regards to long-term synapse plasticity (which are in the range of minutes to years). Due to the polycrystalline nature of HZO, the domain reversal is a stochastic process with a certain distribution of coercive fields. The direct correlation of the ferroelectric properties to the FTJ switching behavior was reported previously using similar devices,[34] where multilevel switching was shown to be stable and reproducible. The capability to gradually change the resistance between multiple states is crucial for emulating synapses. A simple parallel resistor model where each domain is either in the upwards or downwards polarization and therefore in the LRS or HRS can be applied and is schematically shown in figure 1d. The fraction of switched domains can be extracted from polarization current measurements shown in figure 1e. The constant dielectric current contribution has been subtracted to only show the switching current. Cumulative integration of the current gives the fraction of the switched domains, similar to extracting the polarization hysteresis curves. On the other hand, by applying the parallel resistor model, one can calculate the fraction of the switched ferroelectric layer from the overall tunneling electroresistance by the following formula:

$$\frac{1}{R_{total}} = \frac{S}{R_{LRS}} + \frac{1-S}{R_{HRS}}. \qquad (1)$$

Here $R_{total}$ is the overall resistance of the device, $R_{HRS}$ and $R_{LRS}$ are the resistance values of the high resistance and low resistance state respectively, and S is the fraction of the switched domains. In figure 1e, the measured tunneling current is shown as a function of the set voltage amplitude for a fixed pulse width (100μs). Applying the theoretical value from the parallel resistor model and the fraction of switched domains shown with the dotted line in figure 1f, a good agreement can be found for our devices. The parallel resistor model is a suitable model for describing intermediate resistance states. In general, for ferroelectric devices this simple model has been used before, with experimental results that can be observed using PFM techniques.[48]

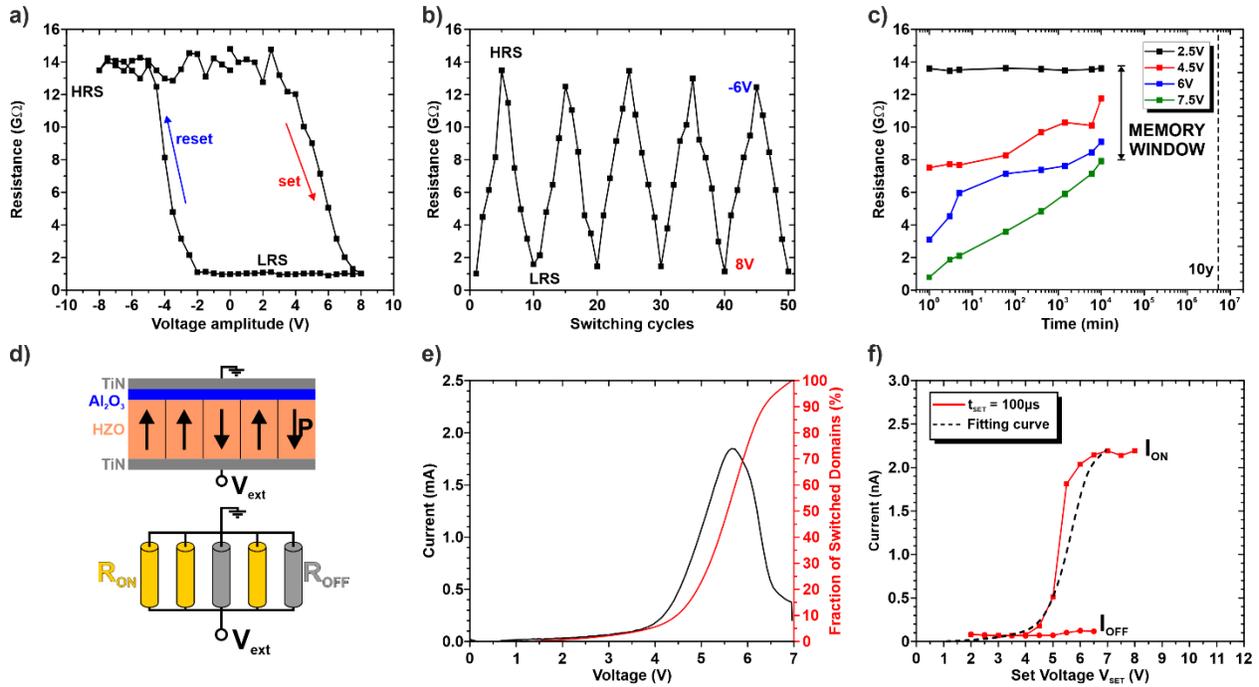

**Figure 1**. a) Resistance hysteresis loop of the ferroelectric tunnel junction as a function of the applied voltage. b) Reproducibility of intermediate resistance states up to 50 pulses resulting in 5 cycles. c) Retention of four different resistance states measured at room temperature. The remaining memory window for the most distant states after $10^4$ minutes is indicated by the double arrow. d) Schematic domain configuration for the MFIM tunnel junction and corresponding parallel resistor model. e) Fraction of switched ferroelectric domains (right axis) extracted from the polarization switching current (left axis). f) Readout current depending on the applied set voltage in comparison to the current expected from the from the parallel resistor model.

For investigation of the switching kinetics, the fraction of switched area from the high to the low resistance state is plotted against the pulse width of the rectangular switching pulse for different pulse heights in **figure 2**. The graph shows the typical voltage-time-trade-off observed for ferroelectric films, where the polarization reversal depends on the duration of the pulse as well as the amplitude. Using the nucleation limited switching (NLS) model as the underlying switching mechanism,[49] the mean switching time can be extracted from the plot in figure 2a. For the NLS model, $t_{mean} \propto \exp(1/V)^n$ can be applied, where $t_{mean}$ is the average switching time, n the exponential parameter and V is the applied voltage. The extracted mean switching time is shown as a function of the pulse amplitude in figure 2b. As expected, the switching is faster for higher applied fields. To achieve switching times between nano- and milliseconds, an external voltage between 5 and 7 V needs to be applied to reach the transition between the LRS and HRS and vice versa. This is in good agreement with previously reported I-V, Q-V and C-V data on similar structures.[50,51] The red fitting curve shows the trade-off between switching time and amplitude, with the best fit for n=1.5 as the exponential parameter for the NLS model. This proportionality and the fitting curve is shown again in figure 2c, with the switching time on a logarithmic scale in dependence of the inverse voltage to the power of 1.5.

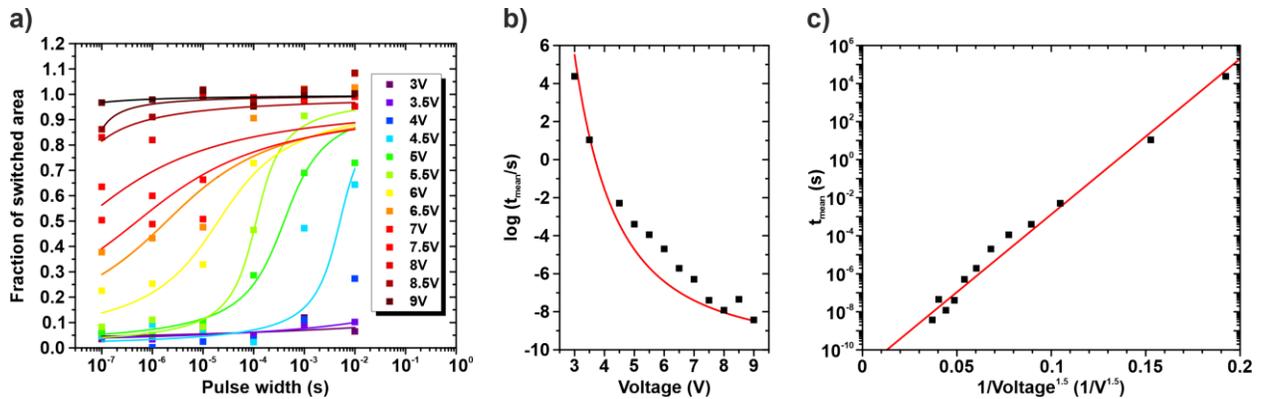

**Figure 2.** a) Normalized fraction of switched area as a function of the pulse time for different pulse amplitudes. b) Extracted logarithmic mean switching time depending on the set voltage. c) Mean switching time as a function of the inverse voltage. A good agreement with the nucleation limited switching model (red line) is observed.

The ability to consistently and reliably tune the resistance is crucial for storing the synaptic weight which interconnects the different nodes in a neural network. In figure 3, the evolution of the TER is shown for different pulsing schemes. Electrical pulses can either have different amplitudes, different pulse widths or can be repeated several times. In figure 3a the evolution of the FTJ resistance is shown when the pulse width is exponentially increased for consecutive pulses (see corresponding scheme in figure 3d), while keeping the pulse amplitude constant at 6 V for the set operation and -3.5 V for the reset operation. Note here that the amplitude needed for set and reset is not symmetrical due to fixed charges at the HZO/aluminum oxide interface[52], leading to built-in bias fields shifting the coercive field of the HZO layer. The transition between LRS and HRS is gradually tunable for both switching directions. The lowest pulse width of 1 µs was only limited by the parasitics of the measurement setup. $Hf_{0.5}Zr_{0.5}O_2$ ferroelectric films in capacitor or FeFET structures have been shown to be switchable at even much shorter voltages pulses,[53,54] which are necessary for realizing semiconductor memories with high write speeds. These results should also hold true for scaled down integrated FTJs to reach the typical timescales in which biological synapses transmit information.[55] Switching between the two resistance states can also be achieved when changing the bias amplitude at the electrodes. This is shown in figure 3b with the pulse scheme depicted in figure 3e. The pulse width was kept constant at 10 µs. With increasing amplitudes on the set pulses, the resistance can be decreased with set voltages higher than 4 V until the LRS is reached at 6 V amplitude, while the inverse switching process with increasing reset amplitudes (absolute values) begins at -3 V and HRS is attained at reset voltages of -5 V. Increasing the voltages further has no switching effect since all ferroelectric domains are already polarized into the opposite direction. In the third pulsing scheme (figure 3f) the amplitude and the pulse width of the bias signal is kept constant and only the number of pulses is varied. As illustrated in figure 3c, after switching the device into the on-state (LRS) by an initial set pulse of 8 V, the resistance can be increased by sequential reset pulses with a fixed amplitude of -3 V and 10 µs pulse width. With each consecutive pulse up to a total of 40, the resistance increases due to accumulated switching effects in the ferroelectric layer.[56] Each voltage stress switches more domains into the downward polarization state, which leads to a nearly linear increase of the resistance levels until the HRS is reached. A similar transition can be found for the set operation. After the initial reset pulse of -5 V the FTJ can be set into the LRS by consecutive set pulses of 4 V and 10 µs. The number of pulses needed for complete switching and therefore the slope of this curve dR/dn (n being the number of pulses) can be adjusted by the amplitude and width of the successive pulses. This is experimentally verified for the reversal from HRS to LRS by consecutive set pulses with 3 different amplitudes of 4 V, 4.5 V and 5 V, while the pulse width is kept constant at 10 µs. For the highest amplitude it takes only a few pulses to switch the device almost completely to the LRS, while for the lowest set voltage a much finer tuning of the resistance value is possible. For all these measurements shown in figure 3 the voltage pulses were only applied at the bottom electrode with the top electrode grounded.

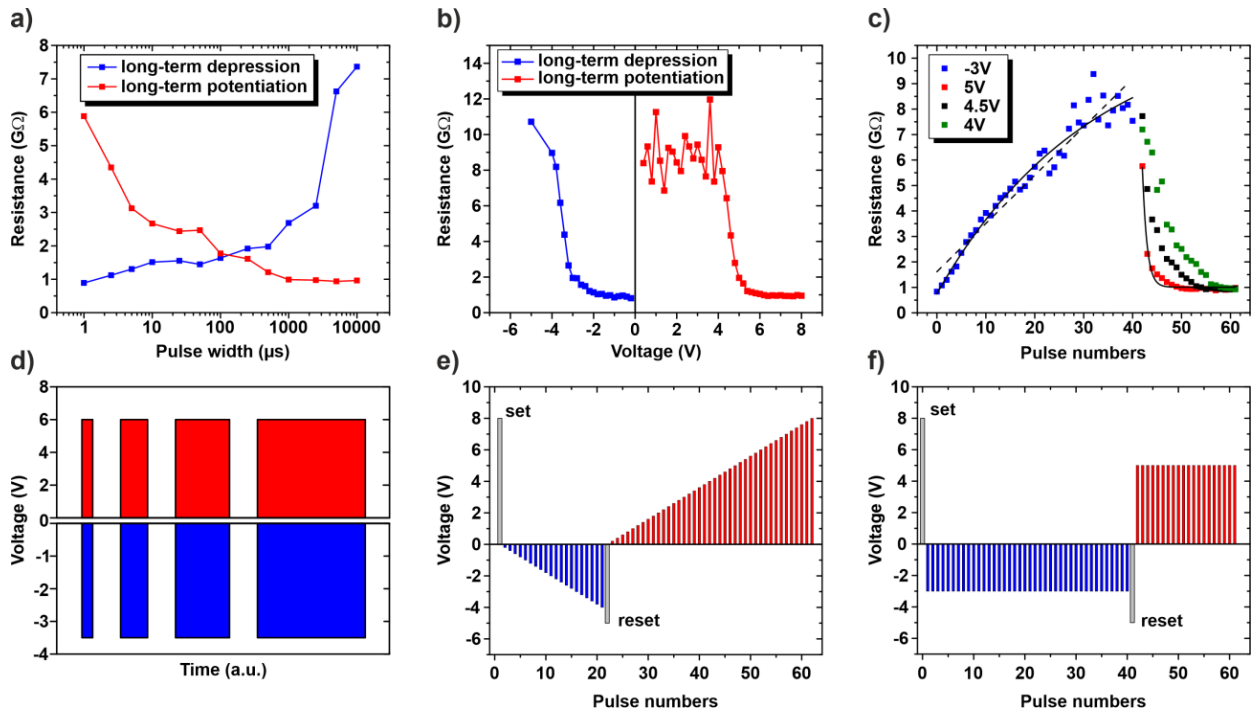

**Figure 3.** Resistance of the ferroelectric tunnel junction as a function of a) pulse width, b) pulse amplitudes, c) number of identical pulses with fixed amplitude and width. The corresponding pulsing schemes are shown in at the bottom d), e) and f) used to create the curves in a), b) and c) respectively.

The transition between HRS and LRS and vice versa was shown to be smooth and gradually tunable for all different pulse schemes by choosing the right pulse width and/or amplitude. This overall behavior is similar to large ferroelectric field effect transistors (FeFETs), which have previously been shown to exhibit similar characteristics that can be used as an artificial synapse.[18] In a FeFET the overall channel resistance is given by the domain configuration of the ferroelectric gate insulator that controls the conducting percolation paths in the channel and therefore the channel resistivity. In ideal FeFETs, no current is transported through the ferroelectric layer and the polarization of the single grains determine the resistance distribution the channel element and eventually the formation of percolation paths. For an FTJ in contrast, the current flows through the ferroelectric layer and the parallel resistor model explained earlier can be applied directly. However, when the device is aggressively scaled down to a size where it only contains a small number of domains (where the device diameter is in the HZO grain size range, e.g. 5-30 nm),[26] we may see a more abrupt switching as it has been observed for scaled down hafnia-based FeFETs.[57,58] More studies will be required to determine if the current transport perpendicular to the ferroelectric layer in FTJs is indeed more favorable in maintaining the continuous switching even for very small feature sizes.

Changing the synaptic weight of the FTJ cell in a spiking neural network through gradual tuning of the resistance state can therefore be used to mimic LTD/LTP behavior, e.g. by assigning the LRS-HRS transition to depression and the HRS-LRS transition to potentiation. This gives significant flexibility of tuning the pulsing sequences generated by the neurons to achieve the required behavior when the FTJ is used as a synapse that connects the neurons in a SNN. In a realistic SNN the effective voltage pulses at the synapse are created by the overlap between the pulses from two neurons connected to the synapses. In this sense spike-timing-dependent-plasticity is an important feature of the synaptic cell. In STDP, the strength of the connection is determined by the timing difference between the action potentials reaching the synapse from both sides. This corresponds to the electrostatic potential to each electrode in our FTJ device. Therefore, we have investigated STDP in our devices. We call the electric signal at the top electrode the pre-neuron action potential $V_{pre}$, and the signal at the bottom electrode the post-neuron action potential $V_{post}$. The schematics of comparing a biological synapse to our TiN/HZO/Al$_2$O$_3$/TiN FTJ can be seen in figure 4. Since the voltages used in the previous measurements were always applied at the bottom TiN layer, the voltage difference $V_{FTJ}$ between the electrodes is here defined as

$$V_{FTJ} = -(V_{pre} - V_{post}). \tag{2}$$

We experimentally emulated the action potential as sawtooth voltage pulses illustrated in figure 4c and 4d, where the black curve represents the pre-neuron signal and the red curve represents the post-neuron signal. The maximum of each half pulse is reached at 4 V after 100 µs. The delay between the voltage signals at pre- and post-neuron is given by Δt. The superposition of both waveforms is the resulting voltage $V_{FTJ}$ from Equation 2 that is illustrated by the blue spiked waveform in figure 4c and 4d. While the individual pulses can only trigger a very small polarization/resistance change, during certain periods of the signal the total voltage across the FTJ becomes high enough to create a significant change of the resistance state of the cell. Therefore, the decrease or increase of the resistance state of the FTJ will depend on the time delay between the pulses. Figure 4b shows the change of the resistance value as a function of the time delay between pre- and post-neuron action potentials, which can be directly interpreted as the change of the synaptic weight of the cell. For small positive time delays, the pre-synaptic action potential precedes the post-neuron action signal, which biologically implies a causality between the information transfer between both ends. The large maximum voltage resulting from the superposition decreases the resistance of the cell (equivalent to set operation into LRS), enhancing the synaptic strength between these two neurons. If Δt becomes too large, no causality between each neuron firing its action potential can be established and no significant change of the synaptic strength occurs. Vice versa, small negative time delays lead to an increase in the resistance state, weakening the synaptic strength when the post-neuron signal fires before the pre-neuron signal.

To quantitatively compare the STDP behavior to other results, the weight change can be fitted by an exponential equation:

$$\Delta R = A_{\pm} \cdot e^{\frac{\Delta t}{\tau_{\pm}}}, \qquad (3)$$

where $A_+$ and $A_-$ represent the maximum scaling factors in the synaptic weight and $\tau_+$ and $\tau_-$ represent time constants that describe how strong the plasticity of the synapse depends on the time delay. For symmetric waveforms for both pre- and post-neuron action potentials as shown in figure 4c and 4d, which correspond to the blue and red lines in the STDP curve (Figure 4b), the extracted parameters are 89 G$\Omega$ and -241 G$\Omega$ for $A_+$ and $A_-$, and -35 µs and 22 µs for $\tau_+$ and $\tau_-$, respectively. Only the time constants $\tau$ are comparable to other STDP curves since the scaling factors A can be arbitrarily changed by shifting the STDP curves to the left or right through different pulse parameters (amplitude and width). The measured time constants are comparable to literature values reported elsewhere for hafnia-based tunnel junctions.[59] To adapt the behavior of the synapse to different requirements, e.g. for different biological systems,[60] the scaling factors and time constants can be adjusted by changing the pulse amplitude and pulse width of the pre- and post-neuron action potentials. When the sawtooth waveform amplitude and pulse width are changed to +6 V and -4 V and 10 µs and 50 µs for the positive and negative branch, respectively, the STDP curves shown in figure 4b in light blue and orange are obtained. The resistance change of the FTJ with regards to the time delay between pulses can therefore conveniently be tuned by choosing appropriate parameters for the excitation signals mimicking the action potentials.

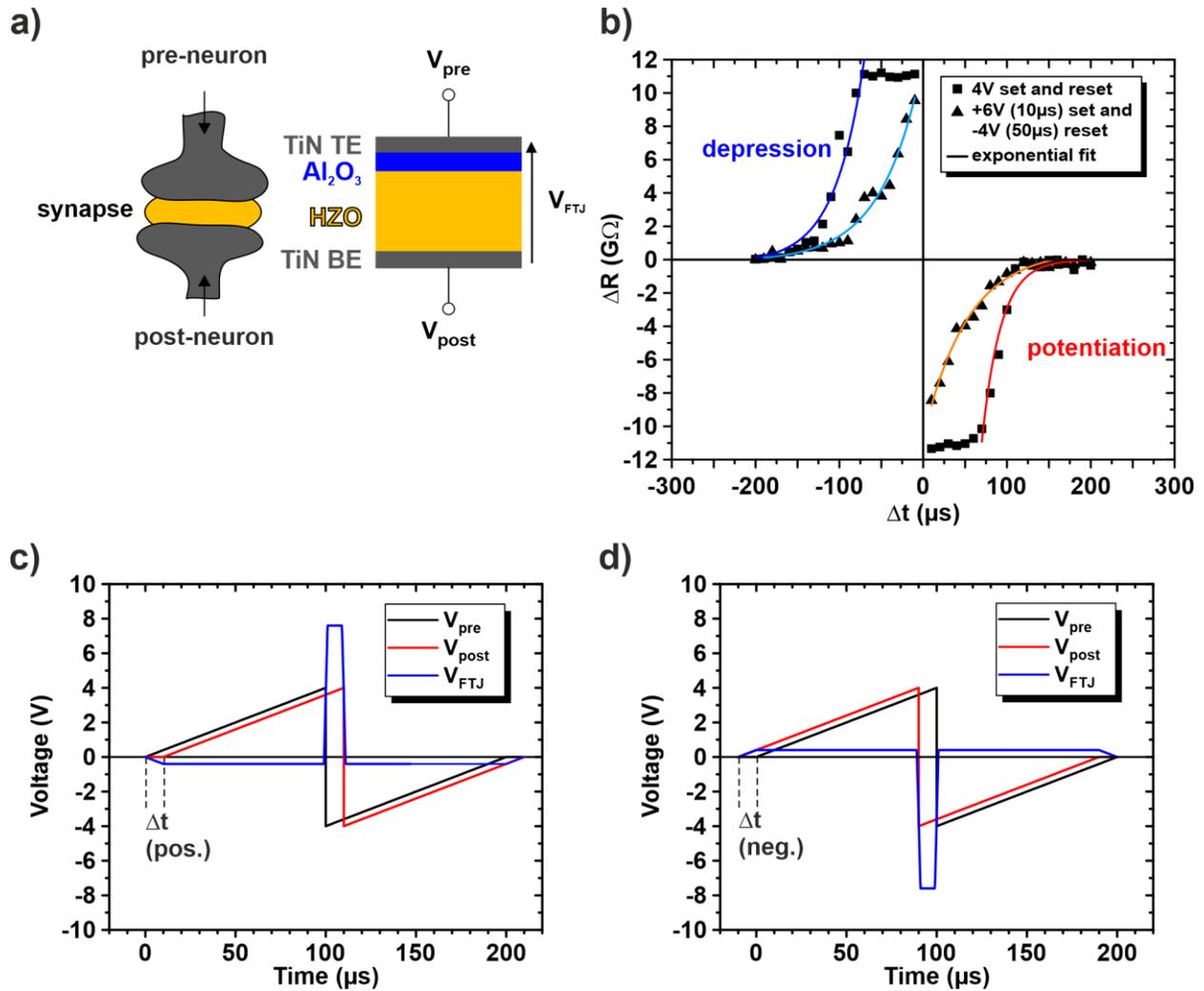

**Figure 4**. a) Schematic comparison between a biological synapse with pre- and post-neuron connection and the FTJ used in this work. b) Measured spike-timing-dependent-plasticity (STDP) curves. The resulting resistance of the FTJ is a function of the time delay Δt between pre- and post-neuron action potentials leading to an increased or decreased resistance of the FTJ that represents a weakening or strengthening of the synapse. c) and d) illustrate the pulse schemes that emulate the action potential created by the pre- and post-neurons with a certain time delay together with the resulting voltage difference created across the FTJ. In c) the time delay between $V_{pre}$ and $V_{post}$ is positive, while in d) the time delay is negative.

By fitting the resistance-voltage curves from figure 1a, the resistivity change of the FTJ when exposed to a given voltage pulse can be predicted. It is therefore possible to emulate different biological systems and their corresponding action potential waveforms for the pre- and post-neurons that might differ from human neurons.[61] The different waveforms are shown in figure 5 with a time delay of +10μs: a triangular voltage pulse with asymmetric amplitudes and pulse lengths that resembles a human action potential more closely (a), a completely rectangular waveform with constant voltage plateaus (b) and a partially rectangular/triangular pulse sequence with opposite signs of pre- and post-neuron (c). The resulting voltage spike across our FTJ structure is given by equation (2) and is shown as the blue curve. The STDP curves predicted from the resistance hysteresis data in figure 1a are shown in figure 5 d-f, where the red spline curve serves as a guide to the eye connection between different resistance change values and gives a good estimation for the complete STDP curve. For the pulse waveform from a), a slight asymmetry in the STDP curve is visible that stems from the non-symmetric and shifted resistance curve due to the fixed charges at the $HZO/Al_2O_3$ interface. For the rectangular pulses a plateau in the STDP curves is visible due to the overall voltage drop being sufficiently high for switching at most time delay values. In c) the pre- and post-neuron action potential have reversed signs, which leads to a high positive set voltage independent of the time delay. Therefore, a symmetric STDP curve is achieved that always leads to a decrease in the resistance and therefore strengthening of the synapse. This symmetric STDP curve has been found in biological systems before.[60] We can show that a broad variety of waveforms with the desired STDP characteristic can be used to emulate the synaptic behavior with our FTJ devices, depending on the specific biological systems that one wants to reproduce.

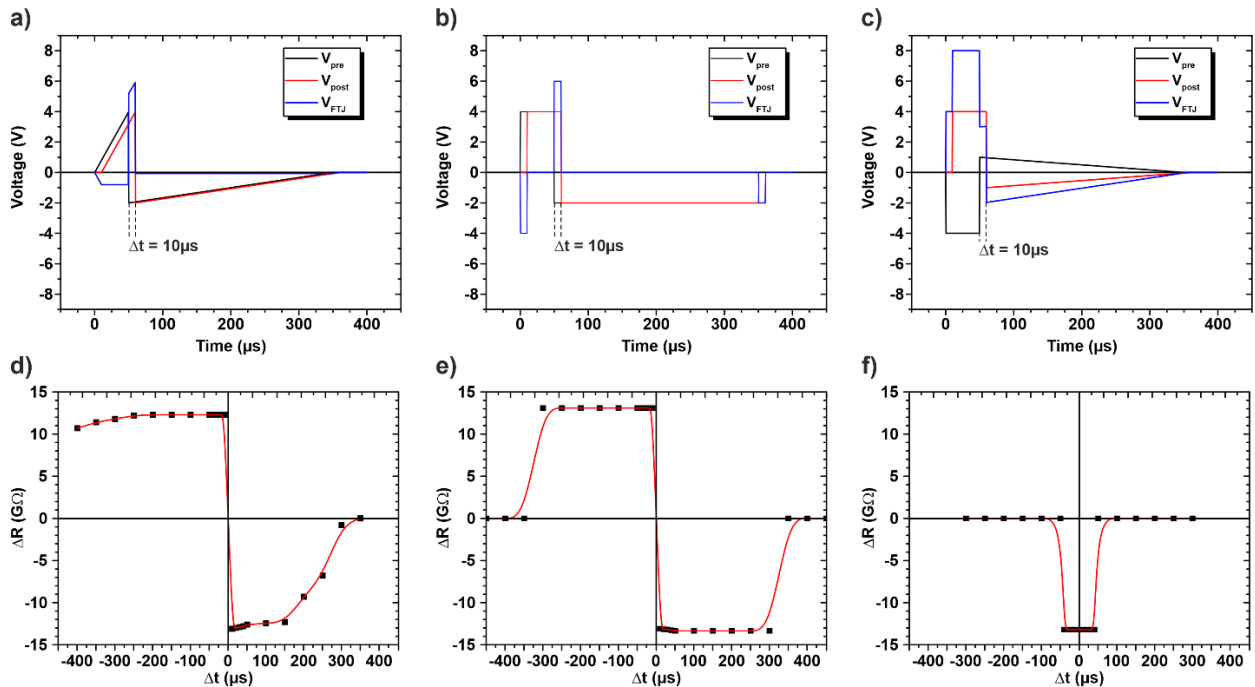

**Figure 5.** a)-c) Different pre- and post-neuron action potential waveforms and the resulting voltage drop across the TiN/HZO/Al$_2$O$_3$/TiN FTJ structure. d)-e) Prediction of the resulting STDP curve and the resistance change following the resistance-voltage-hysteresis shown in figure 1a.

CONCLUSION

In conclusion, we have investigated TiN/Hf$_{0.5}$Zr$_{0.5}$O$_2$/Al$_2$O$_3$/TiN two-layer ferroelectric tunnel junctions with respect to their potential use as artificial synapses. By separating the polarization and the tunneling layer in the FTJ, a stable and efficient ferroelectric switching device is achieved, while an FTJ using only a very thin ferroelectric layer in an MFM structure still suffer from high background leakage currents and reduced ferroelectric properties such as low remanent polarization. Although ultra-thin epitaxial ferroelectrics grown on lattice matched substrates may show better electrical properties like increased current densities or higher memory window compared to the double layer FTJ proposed here,[62,63] there is no viable scenario for integrating such epitaxial layers on lattice matched electrodes into high density neuromorphic circuit in the foreseeable future. Moreover, our approach uses two metal electrodes making a seamless integration into the back end of line of existing technologies possible. To achieve very thin hafnium oxide based ferroelectrics with good polarization, the only published approach also requires to grow the films on a silicon substrate limiting their use to a single layer approach. The tunable tunneling electroresistance with multiple intermediate resistance states of our 2-layer composite FTJs is crucial for the application in spiking neural networks. We have shown that the total resistance can be gradually changed by voltage pulse schemes where either the pulse width (fixed amplitude), pulse amplitude (fixed width) or number of pulses (fixed amplitude and width) is varied. This is the basic mechanism in the long-term depression and long-term potentiation, which in turn are the foundations of learning and memory in the human brain. We have demonstrated that flexible spike-timing-dependent-plasticity can be obtained in the two-layer FTJ. The quantitative analysis of the STDP curves reveals crucial scaling and timing parameters for the learning rate, which can be tuned by the action potential waveform amplitude and width. These results will be beneficial for the implementation of these novel two-layer FTJ devices into future neuromorphic circuits applications. Another attractive property of HZO-based FTJs is the great scaling potential of this material. In fact, fully functional $HfO_2$-based FeFETs having lateral dimensions of only 80 nm and 30 nm for the channel width and length, respectively, have been reported. FTJs are expected to follow a similar scaling trend. Moreover, while the rather low current of the two-layer FTJ is a challenge when using the device in digital memory cells,[64] it is beneficial to realize very large neural networks.[42] Nevertheless, it still remains to be seen whether such nanoscale FTJs will display gradual switching as shown for the larger devices here. More research on this and related issues is necessary in the future.

EXPERIMENTAL

The TiN/Hf$_{0.5}$Zr$_{0.5}$O$_2$/Al$_2$O$_3$/TiN metal-ferroelectric-insulator-metal samples were deposited directly on a p-doped silicon wafer. The TiN bottom electrode with a thickness of 12 nm was sputtered at room temperature using a physical vapor deposition (PVD) process. Atomic layer deposition (ALD) in an Oxford ALD OpAL tool was used to grow the 12 nm HZO layer (1:1 ratio of TEMA-Hf and TEMA-Zr precursor pulses, water as oxygen source) and the 2 nm dielectric aluminum oxide film (TMA and water) at 260 °C directly on the titanium nitride bottom electrode. The 12 nm TiN top electrode was sputtered with the same process as the bottom TiN electrode. A post-deposition anneal of 600 °C for 20 s in nitrogen atmosphere was done to crystallize the HZO layer into the orthorhombic ferroelectric phase. The circular capacitor structure was formed by e-beam evaporation of Ti/Pt dots with 200 μm diameter through a shadow mask and subsequent SC1 etching of the TiN top electrode. The current-voltage response of the samples was analyzed using a Keithley 4200 SCS on a semi-automatic Cascade probe station with pulse measurement units for current and voltage sensing/forcing. Ferroelectric properties were measured by polarization-voltage curves on an Aixacct TF3000 Analyzer, using triangular voltage pulses with a frequency of 10 kHz.

AUTHOR INFORMATION

**Corresponding Author**

Benjamin Max

Email: benjamin.max@tu-dresden.de

**Author Contributions**

The manuscript was written through contributions of all authors. All authors have given approval to the final version of the manuscript.

ACKNOWLEDGMENT

This work was supported by the Free State of Saxony, Germany. S. Slesazeck acknowledges funding by the European Union's Horizon 2020 research and innovation programme under grant agreement No 871737.